Volume 36, Issue 2

# The subjective discount factor and the coefficient of relative risk aversion under time-additive isoelastic expected utility model

Pepin Dominique
*Centre de Recherche sur l''Intégration Economique et Financière*

## Abstract

By analysing the restrictions that ensure the existence of capital market equilibrium, we show that the coefficient of relative risk aversion and the subjective discount factor cannot be high simultaneously as they are supposed to be to make the standard asset pricing consistent with financial stylised facts.

# 1. Introduction

The coefficient of relative risk aversion $\alpha$ and the subjective discount factor $\beta$ are among the most important parameters of human behavior in a risky intertemporal environment. Moreover, under the time-additive isoelastic expected utility model, which is the basic model in asset pricing theory, they are the only parameters available.

What is known about these parameters is that both are supposed to be set high in order to make the standard asset pricing model consistent with financial stylised facts. The subjective discount factor is supposed to be high to justify a low risk-free rate. Estimated values of $\beta$ confirm that is has a high value, « less than, but close to, unity » (Hansen and Singleton, 1983, p. 260). On the other hand, the risk aversion parameter is supposed to be high to explain the historical high risk premium observed on stock markets (see the discussion in Fama, 1991 and Cochrane, 2008). Yet a high $\alpha$ is a dubious assumption (Mehra and Prescott, 1985). Estimates of $\alpha$ in the literature vary widely, but the most frequently estimated values lie between 1 and 3 (Gandelman and Hernández-Murillo, 2015), which is rather low.

Are the subjective discount factor and the risk aversion parameter really high, or not? As we show in this paper, an equilibrium condition drawn from the standard asset pricing model can be used to answer this question. Indeed, for equilibrium to obtain, some restrictions on parameters should be imposed. Surprisingly, the asset pricing literature has so far not been much concerned with these restrictions, which are largely ignored in applied studies. In early studies of asset pricing, authors were a lot more concerned with this problem. Weil (1989, p. 407) mentions that some restrictions have to be imposed but gives no detail. Mehra and Prescott (1985) do provide some details, albeit briefly, on such restrictions, and Mehra (1988) gives a demonstration of the issue for the Mehra-Prescott-type economy, but the existence of the equilibrium conditions is mostly anecdotal in the subsequent huge volume of literature devoted to asset pricing. However, characterizing these restrictions is one of the best ways to inform the debate about the magnitude of $\alpha$ and $\beta$.

# 2. The basic asset pricing equations

Suppose that a single representative household ranks its preferences over their consumption path according to the time-additive expected utility model:

$$E_0\left\{\sum_{t=0}^{\infty}\beta^t U(c_t)\right\}, \beta > 0, \tag{1}$$

where $E_0$ is the expectation operator conditioned on period 0 information, $c_t$ is the per capita consumption, and $U(.)$ is the period utility function.

In a Lucas-type economy, the fundamental pricing relationship is (Lucas, 1978):

$$p_t = E_t\left[\beta\frac{U'(c_{t+1})}{U'(c_t)}(p_{t+1} + c_{t+1})\right], \tag{2}$$

where $p_t$ is the price of the one equity share and $c_t$, the real consumption, is the period dividend.

Let's assume the no-bubble condition. Equation (2) can be solved to get:

$$p_t = \sum_{i=1}^{\infty}E_t\left[\beta^i\frac{U'(c_{t+i})}{U'(c_t)}c_{t+i}\right] \tag{3}$$

Suppose that the utility function is of the isoelastic type:

$$U(c_t) = \frac{c_t^{1-\alpha}}{1-\alpha}, \alpha > 0. \tag{4}$$

Lemma 1. Assume that $x_{t+1} = \ln(c_{t+1}/c_t)$ is an autocorrelated process of the MA(1) type: $x_{t+1} = \mu_x + \varepsilon_{t+1} + \rho\varepsilon_t$, where $\varepsilon_t$ is an iid zero-mean Gaussian innovation with $V(\varepsilon_t) = \sigma^2$. Then the equilibrium asset price is:

$$p_t = a\exp(b\varepsilon_t)c_t, \qquad (5)$$

with $b = (1-\alpha)\rho$ and $a = \dfrac{\exp(k)}{1-\exp(k+e)}$, where $k = \ln\beta + (1-\alpha)\mu_x + \dfrac{(1-\alpha)^2\sigma^2}{2}$ and $e = \dfrac{(1-\alpha)^2(\rho^2 + 2\rho)\sigma^2}{2}$.

Mehra and Prescott (1985) document that the consumption growth rate $x_{t+1}$ is negatively autocorrelated whereas Azeredo (2014) takes the opposite view. If we put $\rho = 0$ in the above equations, we recover the standard equation of Mehra (2003) and Mehra and Prescott (2003) when $x_{t+1}$ is iid. In this case, another important result of the model is the standard risk premium equation:

$$\ln E(R_{e,t+1}) - \ln R_{f,t+1} = \alpha\sigma_x^2 \qquad (6)$$

where $R_{e,t+1} = (p_{t+1} + c_{t+1})/p_t$ is the gross return on equity, $R_{f,t+1}$ is the risk-free rate and $\sigma_x^2 = V(\ln x_{t+1})$.

It follows from equation (6) that the equity premium is the product of the coefficient of relative risk aversion and the variance of the growth rate of consumption. The model seems inconsistent with the facts that $\sigma_x^2$ is very low and the equity premium is high unless $\alpha$ is very large (Mehra and Prescott, 1985).

### 3. The existence of equilibrium asset prices

Equilibrium asset prices equal expected discounted dividends. But for equilibrium asset prices to be defined, the series defined by the right-hand side of equation (3) must converge. The restrictions that ensure convergence of this series are the same that guarantee convergence of the expected utility (1). It is well-known that the existence of the expected isoelastic utility is fragile and that some distributional assumptions are more consistent with it (Geweke, 2001 and Yoon, 2003). But the existence of expected utility is also fragile with respect to changes in the preference ($\alpha$ and $\beta$) and the technological ($\mu_x$ and $\sigma_x^2$) parameters. In the case of the simple economy described in this paper, the conditions of existence of equilibrium are straightforward to derive, because the asset price is given by equation (5) if equilibrium does obtain. Then the condition of convergence of the series (1) and (3) is simply the condition of existence of a positive and finite $p_t$ in (5), defined by $a > 0$. Clearly, this condition is the same as the restriction $k + e < 0$. By noting that $\sigma_x^2 = (1+\rho^2)\sigma^2$, this restriction can be written as $\ln\beta + (1-\alpha)\mu_x + \dfrac{(1-\alpha)^2\sigma_x^2}{2} - \dfrac{\rho(1-\alpha)^2\sigma_x^2}{1+\rho^2} < 0$.

Theorem 1. Under the assumptions of lemma 1, the existence of equilibrium requires that:

$$\beta < \beta' = \exp\left[-(1-\alpha)\mu_x - \frac{(1-\alpha)^2}{2}\sigma_x^2\right]\exp\left[-\frac{\rho(1-\alpha)^2\sigma_x^2}{1+\rho^2}\right] \qquad (7)$$

Suppose that the value of the technological parameters $\mu_x$ and $\sigma_x^2$ are fixed. Equation (7) states that, for each value of α, there is a maximal value β' that β must not exceed in order to ensure existence of equilibrium. Suppose that $\sigma_x^2 = 0.00125$ and $\mu_x = 0.0172$ (Mehra and

Prescott, 1985 and 2003). Table 1 presents the function β' evaluated at $\sigma_x^2 = 0.00125$ and $\mu_x = 0.0172$ for different values of ρ.

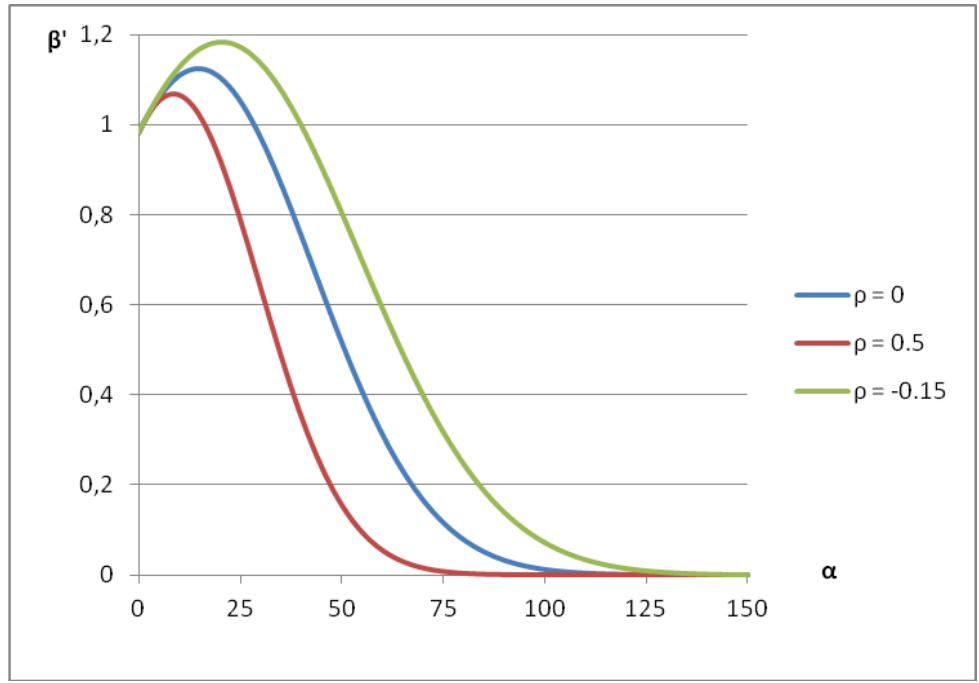

**Graph 1.** β' as a function of α for different values of ρ

Inspection of graph 1 shows that β' is sensitive to the value of ρ. We represent the function β' for three different values of ρ. The lowest value, ρ = -0.15, roughly corresponds to a -0.14 autocorrelation, which is the negative autocorrelation of consumption growth documented by Mehra and Prescott (1985). The intermediate value, ρ = 0, is that of an iid process. The highest value, ρ = 0.5, corresponds to a 0.4 autocorrelation, which is roughly the level of positive autocorrelation documented by Azeredo (2014). Whatever the value of ρ is, graph 1 shows that for a high value of α, β' is low. As a result, α and β cannot be high simultaneously. Moreover, the results indicate that the higher the coefficient ρ is, the lower β' is. If we consider the standard case ρ = 0, and fix $\alpha = 55$, which is the value of risk aversion needed to fit the risk premium equation according to some authors (Mehra and Prescott, 1985 and Cochrane, 2008), equation (7) yields β'= 0.409, which is significantly lower than the most frequently estimated values of β. For ρ = 0.5 and $\alpha = 55$, the maximal value of the subjective discount factor consistent with equilibrium is β'= 0.095, which is extremely low.

## 4. Discussion and conclusion

Azeredo (2014) is an important reference, defending the view that raising risk aversion leads to higher risk premium. He claims that the equity premium turns negative for moderately high value of the risk aversion. But he gets to such a result by assuming that the consumption autocorrelation is positive. Another problem with the standard model of asset pricing we point out in this paper is that the coefficient of relative risk aversion cannot be high if we assume that the subjective discounted factor is high too.

If both parameter values are high, equilibrium does not obtain. Practically, this means that, even if $\rho = 0$, the risk premium equation (6) is invalid. Suppose for example that

$\rho = 0, \beta = 0.9$ and $\alpha = 30$, and use equation (6) to calculate the theoretical risk premium $\ln E(R_{e,t+1}) - \ln R_{f,t+1} = 0.0375$. Then it may be tempting to conclude that for $\rho = 0, \beta = 0.9$ and $\alpha = 30$ the model predicts a risk premium of 3.75 percent; but in fact this is a misleading result because if we calculate the stock price from equation (5) we get a negative value as $a = -1.033$.

The first conclusion to be drawn is that a high coefficient of relative risk aversion does not necessarily constitute a solution to the risk premium puzzle documented by Mehra and Prescott (1985), which exacerbates the puzzle.

The second conclusion to draw is that researchers in asset pricing theory have to be very cautious with the numerical implementation of their models. The standard model discussed in this paper is the building block of more complex models used nowadays. These models are solved by numerical methods, looking at the returns on assets. But as shown above the simulation of a model can give seemingly consistent results in terms of returns, whereas in fact there is no equilibrium in asset prices.

## Appendix

**Proof of Lemma 1**

We hypothesize that the solution of (2) has the form $p_t = a \exp(b \varepsilon_t) c_t$. By inserting this general solution into (2), we get:

$$a \exp(b \varepsilon_t) = E_t \left[ \beta \left( \frac{c_{t+1}}{c_t} \right)^{1-\alpha} a \exp(b \varepsilon_{t+1}) + \beta \left( \frac{c_{t+1}}{c_t} \right)^{1-\alpha} \right]$$

$$= a E_t [\exp(\ln \beta + (1-\alpha) x_{t+1} + b \varepsilon_{t+1})] + E_t [\exp(\ln \beta + (1-\alpha) x_{t+1})]$$

If $z \sim N(\mu_z, \sigma_z)$, then $E[\exp(z)] = \exp\left[\mu_z + \frac{1}{2} \sigma_z^2\right]$. Hence:

$$a \exp(b \varepsilon_t) = a E_t \left[ \exp\left( \ln \beta + (1-\alpha)(\mu_x + \rho \varepsilon_t) + \frac{(1-\alpha)^2 + b^2 + 2(1-\alpha)b}{2} \sigma^2 \right) \right]$$

$$+ E_t \left[ \exp\left( \ln \beta + (1-\alpha)(\mu_x + \rho \varepsilon_t) + \frac{(1-\alpha)^2}{2} \sigma^2 \right) \right]$$

After some rearrangements, we can write this equation as:
$a \exp(b \varepsilon_t) = \exp(k)[1 + a \exp(e)] \exp[(1-\alpha) \rho \varepsilon_t]$

From this equation, a and b are easily identified as:

$b = (1-\alpha)\rho$ and $a = \dfrac{\exp(k)}{1 - \exp(k+e)}$.